\documentclass[pre,aps,amsmath,amsfonts,amssymb,twocolumn]{revtex4-2}
\usepackage{xcolor}
\usepackage{graphicx,CJK}
\begin{document}
\title{Finite-size scaling of the Kuramoto model at criticality}
\begin{CJK*}{UTF8}{mj}
\author{Su-Chan Park (박수찬)}
\email{spark0@catholic.ac.kr}
\affiliation{Department of Physics, The Catholic University of Korea, Bucheon 14662, Republic of Korea}
\author{Hyunggyu Park (박형규)}
\email{hgpark@kias.re.kr}
\affiliation{Quantum Universe Center, Korea Institute for Advanced Study, Seoul 02495, Republic of Korea}
\begin{abstract}
The asymptotic scaling behavior of the Kuramoto model with finite populations has been notably elusive, despite comprehensive investigations
employing both analytical and numerical methods. In this paper, we explore the Kuramoto model with  ``deterministic'' sampling of
natural frequencies, employing extensive numerical simulations and reporting the asymptotic values of the finite-size scaling exponents,
which deviate significantly from the previously reported values in the literature. Additionally, we observe that these exponents are sensitive to the specifics of the sampling method. We discuss the origins of this variability through the self-consistent theory of entrained oscillators.
\end{abstract}
\date{\today}
\maketitle
\end{CJK*}

\section{\label{Sec:intro}Introduction}
The Kuramoto model~\cite{BookK1984}, a prototypical model of synchronization, is a system of
$N$ interacting phase oscillators that evolve in a deterministic manner by the equation
\begin{align}
\frac{d\theta_k}{dt} = \omega_k + \frac{K}{N} \sum_{j\in I} \sin(\theta_j-\theta_k),
\label{Eq:kura}
\end{align}
where $I$ is an index set with $N$ elements, $k$ is any element of $I$,
$\theta_k$ is the phase of an oscillator indexed by $k$
with a corresponding (time-independent) natural frequency
$\omega_k$ that can be any real number, and $K$ is the interaction parameter that is assumed to be positive
(for a review, see, e.g, Ref.~\cite{S2000}).
The natural frequencies are sampled in such a way 
that under the infinite-$N$ limit they should satisfy
\begin{align}
\lim_{N\rightarrow\infty} \frac{1}{N}\sum_{k\in I}\Theta(\omega_k - a) \Theta(b-\omega_k) = \int_a^b g(\omega) d\omega,
\label{Eq:omega_cond}
\end{align}
for any real numbers $a$ and $b$ with $a \le b$, where
$\Theta(x)$ is the Heaviside step function and
$g(\omega)$ is a (normalized) frequency density function that also stipulates the Kuramoto model.
In this paper, we limit ourselves to the mostly studied case that
$g(\omega)$ is a continuous, symmetric, and unimodal function
whose shape near $\omega\approx 0$ is concave and parabolic.

It is customary to introduce the phase order parameters
\begin{align}\label{Eq:order0}
\Delta_N(t):= \frac1N \sum_{j\in I} e^{i\theta_j(t)},\quad
r_N(t) := \left \vert \Delta_N(t) \right \vert,
\end{align}
and to rewrite Eq.~\eqref{Eq:kura} as
\begin{align}
\frac{d\theta_k}{dt}
=\omega_k - K r_N(t) \sin \left [ \theta_k-\psi(t) \right ],
\label{Eq:kuraO}
\end{align}
where the average phase angle, $\psi(t)$, is obtained from
$\Delta_N(t) \equiv r_N(t) e^{i\psi(t)}$.
The long-time behavior of the order parameters are defined as
\begin{align}
\label{Eq:od_OP}
R_N := \lim_{t\rightarrow\infty} \overline{r_N}(t), \quad
R := \lim_{N\rightarrow\infty} R_N,
\end{align}
where $ \overline{X}(t)$ denotes the time average of $X$, defined as $ \overline{X}(t):= \frac{1}t \int_0^t X(t') dt'.$
It is well known~\cite{BookK1984} that
there exists a synchronization transition such that $R=0$ for $K \le K_c = 2/[\pi g(0)]$ and $R \sim (K-K_c)^{\beta}$ for $K>K_c$
with the order parameter exponent $\beta=\frac12$.
 
The dynamic fluctuations of the order parameter can be  defined as
\begin{align}
\label{Eq:fluctuation}
\chi_N := N \lim_{t\rightarrow\infty}\left [ \overline{r_N^2}(t) - \overline{r_N}(t)^2  \right ],\quad
\chi := \lim_{N\rightarrow\infty} \chi_N,
\end{align}
which are expected to behave as
$\chi \sim |K-K_c|^{-\gamma}$ near the transition ($0<|K-K_c|\ll 1$) with
the fluctuation exponent $\gamma$.

For sufficiently small $\epsilon := |K-K_c|$ and for sufficiently large $N$, 
the leading behavior of $R_N$ is expected as
\begin{align}
R_N \sim \begin{cases}
\epsilon^\beta \rho_+\!\!\left (\epsilon^{\bar \nu_+} N \right ),&\epsilon > 0,\\
\\
|\epsilon|^\beta \rho_-\!\!\left (|\epsilon|^{\bar \nu_-} N \right ),&\epsilon < 0,\\
\\
N^{-\beta/\bar \nu_c}, & \epsilon=0,
\end{cases}
\label{Eq:Rchi_sf}
\end{align}
where $\rho_\pm$ are scaling functions that should satisfy
\begin{align*}
\lim_{x\rightarrow\infty} \rho_+(x) = \text{const},&\quad
\lim_{x\rightarrow\infty} \rho_-(x) = 0,
\end{align*}
to conform to the critical behavior of $R$.
In what follows, we will refer to $\bar \nu_+$, $\bar \nu_-$, and
$\bar \nu_c$ as the finite-size scaling (FSS) exponents for
the supercritical region, for the subcritical region, and at the critical point, respectively.
Needless to say, the infinite $x$ limit above should be understood as the infinite $N$ limit
for fixed small $\epsilon$.

If the conventional theory of critical phenomena is applicable to the Kuramoto 
model, one can claim $\bar \nu_+= \bar \nu_- = \bar \nu_c$.
Accordingly, there is no need to distinguish $\bar \nu$'s in Eq.~\eqref{Eq:Rchi_sf}
and the three exponents are usually denoted by a single exponent $\bar \nu$.
In this case, it is a common numerical practice to 
estimate $\bar \nu$ by studying $R_N$ only at the critical point.
As we will see from Sec.~\ref{Sec:SC} on, however, the three FSS exponents of the Kuramoto
model may have different values at the corresponding regimes, depending on the sampling
rule of natural frequencies. On this account, 
we will stick to using the three exponents for the FSS of $R_N$.
Only when we refer to the exponent in the literature, 
we will use $\bar \nu$.
Further discussion about the FSS exponents can be found in Sec.~\ref{Sec:nuc}.

Again by the conventional theory, one may expect the hyperscaling relation
\begin{align}
\label{Eq:hyper}
\gamma=\bar \nu - 2\beta,
\end{align}
which should be correct 
unless the leading finite-size term in $\overline{r_N^2}$ and
$\overline{r_N}^2$ exactly cancel each other in Eq.~\eqref{Eq:fluctuation}.

It is well known that finite-size effects are influenced by the sampling methods of 
natural frequencies (see also Refs.~\cite{P2005,OLS2016} for the same phenomenon
in the uniformly distributed $g$). 
In particular, the values of the exponents
$\bar\nu$ and $\gamma$ 
may also vary depending on the specific details of samplings.
In the literature, two typical samplings 
called {\em random} and {\em regular} samplings
have been under consideration.
In the random-sampling case,
$\omega_k$'s are independent and identically distributed (iid) random variables
in accordance with the density $g(\omega)$.
In the regular-sampling case, $\omega_k$ is determined by
\begin{align}\label{Eq:regular}
\int_{-\infty}^{\omega_k} g(\omega) d\omega = \frac{k-1/2}{N},
\end{align}
with integer $k=1,2,\ldots,N$, comprising the index set $I$.
In Ref.~\cite{HCTP2015}, it was claimed that $\bar \nu = 5/2$ and $\gamma=1$ for
the random-sampling case, and $\bar \nu \simeq 5/4$ and $\gamma \simeq 1/4$ for the regular-sampling case.
For the random-sampling case, these exponent values are well established through some analytic
derivations~\cite{D1986,HBC2007,Chow07,Hong07}
reinforced by extensive numerical simulations~\cite{Hong05,Hong07,HCTP2015}. 
However, for the regular-sampling case,
the exponent values are estimated mostly through numerical simulations for rather small size
and short simulation time~\cite{HCTP2015}. In this paper, we perform extensive numerical
simulations for much bigger size and much longer simulation time to estimate
the accurate exponent values in the asymptotic (large $N$) regime for more general 
``deterministic'' sampling including the regular case.
The main focus in this paper will be on the scaling behavior of $R_N$, or, equivalently, on the value
of $\bar \nu$ (more precisely, $\bar \nu$'s).

The structure of the paper is as follows. 
In Sec.~\ref{Sec:sampling}, we specify the form of $g(\omega)$
as well as the sampling method.
In Sec.~\ref{Sec:Meth}, we explain
details of the method in numerical analysis.
In Sec.~\ref{Sec:sh}, we present our numerical results for the regular-sampling case, which
will be called the equally spaced (ES) case later.
In particular, we will claim that our estimate of $\bar \nu_c$ is actually different from 
$\bar \nu$ in Ref.~\cite{HCTP2015}.
To grasp some analytic understanding, we study the mean-field self-consistent equation (MFSCE)
in Sec.~\ref{Sec:SC}. In Sec.~\ref{Sec:num}, we present numerical results for other sampling 
cases. In particular, we will claim another value of $\bar \nu_c$.
These various critical exponents are discussed in more detail in Sec.~\ref{Sec:dis}.
We summarize and conclude the paper in Sec.~\ref{Sec:sum}.

\section{\label{Sec:sampling}Frequency samplings}
\begin{figure}
\includegraphics[width=\linewidth]{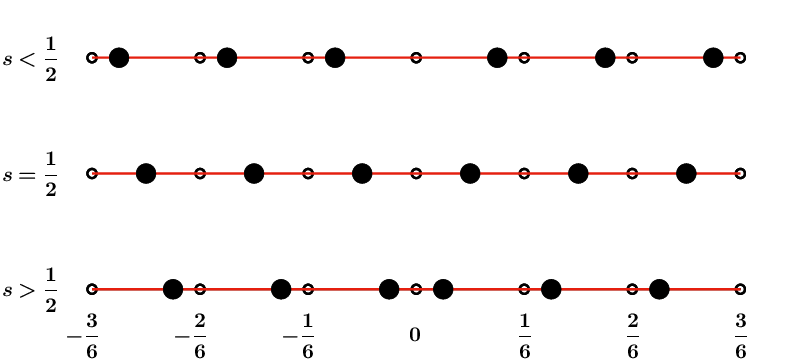}
\caption{\label{Fig:xk}
Example of $x_k$'s with $N=6$ for $s=\frac14<\frac12$, $\frac12$, and $\frac34>\frac12$ (top to bottom).
Filled circles indicate $x_k$'s of the corresponding $s$.
}
\end{figure}
In the case that $g(\omega)$ has an unbounded support, 
a certain oscillator can assume a very high frequency particularly when
$N$ is large.
For example, if $g(\omega) = \pi^{-1}/(\omega^2+1)$ and $N=2^{18}$ (the largest
system size we used in our numerical studies),
the largest frequency in the regular sampling is obtained, by 
setting $k = N$ in Eq.~\eqref{Eq:regular} and using
$$
\frac{1}{\pi}\int_{\omega_k}^{\infty} \frac{d\omega}{\omega^2+1}
=\frac1{\pi} \cot^{-1}(\omega_k),
$$
as $\cot(\pi/2^{19}) > 10^5$, 
which is hard to be controlled in numerical integrations. Although no serious problem due to
this difficulty has been reported, we would like to avoid any unexpected numerical uncertainty
by too fast oscillations.

As the critical scaling only depends on the
shape of $g(\omega)$ near $\omega\approx 0$ and overall unimodality, 
it would be numerically practical to consider a frequency density with a compact support.
Specifically, we consider 
\begin{align}
\label{Eq:gden}
g(\omega) = 
\frac{3}{2} \left ( 1 - 4 \omega^2 \right )\Theta(1-2|\omega|),
\end{align}
which gives $K_c = 4/(3\pi)\approx 0.4244$.

A set of $N$ natural frequencies $\{\omega_k\}$ is sampled deterministically in the following way.
For even $N=2M$, it is convenient to take the index set as $I=\{\pm 1, \pm 2, \ldots, \pm M\}$.
To assign the frequency value $\omega_k$ for each index $k$, we first introduce
\begin{align}
\label{Eq:defxk}
x_k = \begin{cases}
\left ( k-s \right )/N, & k>0 ,\\
-x_{|k|}, & k<0,
\end{cases}
\end{align}
with a constant $s$ in the range $0 \le s \le 1$ (see Fig.\ref{Fig:xk}).
The $\omega_k$ is then determined through the relation
\begin{align}\label{Eq:Gomegak}
G(\omega_k)=x_k
\end{align}
with the function $G$ defined by
\begin{align}\label{Eq:Gomega}
G(\omega) := \int_0^\omega g(\omega') d\omega',\quad\textrm{for}\quad \omega\ge 0.
\end{align}
As $g(-\omega) = g(\omega)$, $G$ is naturally extended to all real numbers by $G(-\omega) = - G(\omega)$.
Equation~\eqref{Eq:Gomegak} can be inverted analytically with $g(\omega)$ given by Eq.~\eqref{Eq:gden}.
Since $G(\omega) = \frac12 (3\omega - 4 \omega^3)$ 
and $\sin (3z) = 3 \sin(z) - 4 \sin^3(z)$,
setting $\omega_k = \sin(z)$ gives $\sin(3z) = 2x_k$ and, therefore,
$$
\omega_k = G^{-1}(x_k)=\sin \left ( \frac{1}{3} \sin^{-1}  ( 2x_k) \right ).
$$

Through this type of sampling, we maintain
the natural frequency symmetry even for finite $N$; $\omega_{-k}=-\omega_k$.
The $s=\frac12$ case corresponds to the regular-sampling case in Eq.~\eqref{Eq:regular}.
We will refer to this case as the ES case (equal interval between any two 
$x_k$'s in a row).
Deviations from $s=\frac12$ govern the non-uniformity of frequency distribution at the finite-size level
near $\omega\approx 0$ (either thinning for $s<\frac12$ or thickening for $s>\frac12$; see Fig.\ref{Fig:xk}).

For odd $N=2M+1$, we take $I=\{0, \pm 1, \pm 2, \pm M\}$ with
$x_0=0$, $x_k=(k-s+\frac12)/N$ for $k>0$, and $x_{-k}=-x_{k}$.
The $s=\frac12$ case also corresponds to the ES case.
The following conclusion for even $N$ does not change for odd $N$ as long as this rule with the same $s$ 
is employed.

\section{\label{Sec:Meth}Numerical methods}

Exploring the FSS behavior near the criticality, especially for deterministic sampling of
natural frequencies, is exceedingly challenging, due to huge computational costs.
In previous studies for the regular sampling, simulations are limited
up to a system size of $N=25~600$ and a simulation time of $t_\text{max}\approx 10^5$~\cite{HCTP2015}.

In our research, we extend these limits considerably to a much larger scale
($N=262~144$ and $t_\text{max}\approx 10^9$) to reach the asymptotic scaling regime.
Moreover, we enhance the reliability of our estimates for the FSS exponent values
by employing an ``effective-exponent'' analysis of extensive numerical data, which
incorporates corrections to scaling.

For convenience, the system size $N$ is assumed even with $N=2M$.
Since $\omega_{-k}= -\omega_k$, it is convenient to rewrite the evolution equations with new variables $\phi_k := \frac12 (\theta_k - \theta_{-k})$
and $\sigma_k := \frac12 (\theta_k + \theta_{-k})$ as
\begin{align*}
\frac{d\phi_k}{dt} &= 
\frac{\omega_k - \omega_{-k}}{2} - K r_N \frac{\sin(\theta_k - \psi) - \sin(\theta_{-k} - \psi)}{2}\\
&=\omega_k - K r_N \cos(\sigma_k -\psi) \sin(\phi_k),\\
\frac{d\sigma_k}{dt} 
&=\frac{\omega_k + \omega_{-k}}{2} - K r_N \frac{\sin(\theta_k - \psi) + \sin(\theta_{-k} - \psi)}{2}\\
&= - K r_N \sin(\sigma_k -\psi) \cos(\phi_k),
\end{align*}
where $k=1,2,\ldots, M$.
From the above equation, one can easily deduce that
if we adopt a ``symmetric'' initial configuration with 
$\sigma_k(0) = \psi(0)=\psi_0$ (const) for all $k$, then
$\sigma_k(t) = \sigma_k(0)$ for all $t$.
Without losing generality, we can set $\psi_0 = 0$ and we have the
symmetry $\theta_{-k}(t) = - \theta_k(t)$ for all $k$ and $t$.

As long as 
$
\lim_{N\rightarrow\infty} \ln R_N/\ln N
$
at criticality is well-defined (with the value $-\beta/\bar \nu_c$),
we expect that an initial condition would not affect the value of $\bar \nu_c$.
In this regard, we choose the symmetric initial configuration that gives
$\theta_k = \phi_k = -\theta_{-k}$ as well as $\sigma_k(t) = 0$
and rewrite the equation with rescaled time $\tau := Kt$ as ($1 \le j \le M$)
\begin{equation}\label{Eq:Rkura}
\begin{aligned}
\Delta_N &= \frac{1}{N} \sum_{k=1}^M \left ( e^{i \theta_k} + e^{i \theta_{-k}} \right )
= \frac{1}M\sum_{k=1}^M\cos\theta_k ,\\
\frac{d\theta_j}{d\tau}&= 
\nu_j - r_N\cos\psi \sin \theta_j
=\nu_j - \Delta_N \sin \theta_j,
\end{aligned}
\end{equation}
where we have used rescaled frequencies $\nu_j:= \omega_j/K$.
Obviously, $r_N \cos \psi = \Delta_N$ because $\Delta_N$ is real.
This equation obviously reduces computational efforts. 
With this symmetric initial condition, the order parameter $\Delta_N$ becomes always real 
and the average phase angle $\psi(t)$ assumes only zero or $\pi$, corresponding
to positive or negative $\Delta_N(\tau)$, respectively.
We confirmed numerically that alternative initial conditions with broken symmetry (for example,
random $\phi_k$ and $\sigma_k$ at $t=0$)
also converge to the same values of the FSS exponents (not shown here).

For numerical integration of Eq.~\eqref{Eq:Rkura},
we employ the fourth order Runge-Kutta method with time step size $d\tau = 0.05$.
As the initial conditions,  we set $\theta_k(0)=0$ for all $k\in I$ in most cases.
To ascertain the stationary-state values of the order parameters,
we disregard the data from the transient regime up to $\tau=\tau_s$ and
collect the data over the interval $\tau_s<\tau \le \tau_s + \tau_\text{max}$
with $\tau_\text{max} = 5 \times 10^8$ (in terms of the original time, the
maximum time is about $t_\text{max} \approx 10^9$).
We set $\tau_s=10^3 N^{2/3}$ to ensure the system has reached a stationary state,
which is supported both numerically
and through theoretical analysis detailed in Sec.~\ref{Sec:num}.

In the stationary state, we calculate the time average of $r_N(\tau)$ as
\begin{align}
\label{Eq:RNtau}
\widetilde R_N(\tau) = \frac{1}{\tau}\sum_{i=1}^\tau  r_N(i+\tau_s),
\end{align}
where $1 \le \tau \le \tau_\text{max}$. Note that data sampling
in taking time average does not occur at every time step, but rather at
every integer value of $\tau$ to reduce correlations between neighboring data points.
Its statistical error is determined by the difference between
the maximum and the minimum of the set
$
\left \{ \widetilde R_N(\tau)
: 3\times 10^8 \le \tau \le 5\times 10^8  \right \}.
$
To support numerical reliability, 
numerical integrations 
with a larger step size $d\tau=0.1$ 
were also performed,
to observe no significant differences in the time averaged quantities (not shown here).

We consider various system sizes, specifically $N = 2^5, 2^{6}, \ldots, 2^{18}$
for  $s=0$, $\frac14$, $\frac12$, $\frac34$, and $1$.
Our primary focus is on the critical FSS at $K=K_c=4/(3\pi)$, where an
averaged quantity $R_N$ is expected to behave for large $N$,
according to the scaling ansatz \eqref{Eq:Rchi_sf},
as
\begin{align}
\label{Eq:Xi}
R_N = a N^{-\beta/\bar\nu_c} \left [ 1 + B N^{-c_R} + o(N^{-c_R}) \right ],
\end{align}
with the leading correction-to-scaling (CTS) exponent $c_R$ that is to be
determined,
where $a$ and $B$ are constants and $o[f(N)]$ denotes higher-order terms beyond $f(N)$.
The crucial assumption that the leading correction to scaling is of the power-law type 
is only supported \textit{a posteriori} by numerical studies.

With a sequence of $R_N$ values for various $N$, we determine $-\beta/\bar \nu_c$ as well as $c_R$ in the
following manner.
For a positive $b>1$, the effective exponent $\xi^\text{eff}$ is defined as
\begin{align}\label{Eq:effective1}
\xi^\text{eff}(N;b) := \frac{\ln R_N -\ln R_{N/b}}{\ln b}.
\end{align}
Since 
\begin{align*}
\frac{R_N}{R_{N/b}} &= b^{-\beta/\nu_c}\frac 
{1 + B N^{-c_R} + o(N^{-c_R})}
{1 + B (N/b)^{-c_R} + o(N^{-c_R})}\\
&=b^{-\beta/\nu_c} \left [ 1 + B(1-b^{c_R}) N^{-c_R} + o(N^{-c_R}) \right ],
\end{align*}
the effective exponent behaves for large $N$ as
\begin{align}
\xi^\text{eff}(N;b) 
= -\frac{\beta}{\bar \nu_c} - B \frac{b^{c_R} -1}{\ln b} N^{-{c_R}} + o(N^{-{c_R}}).
\label{Eq:effective2}
\end{align}
Plotting $\xi^\text{eff}$ as a function of $N^{-{c_R}}$ results in a straight line
for large $N$, enabling the estimation of $-\beta/\bar \nu_c$ via simple linear extrapolation.
Therefore, accurate information regarding the CTS exponent $c_R$ is essential for
obtaining a precise estimate of $-\beta/\bar \nu_c$.
Since $R_N$ is numerically estimated as 
$R_N^- \le R_N \le R_N^+$ ($R_N^\pm$ are numerical values),
the error of $\xi^\text{eff}$ is estimated by
$$
\frac{\ln R_N^- - \ln R_{N/b}^+}{\ln b} \le \xi^\text{eff}
\le \frac{\ln R_N^+ - \ln R_{N/b}^-}{\ln b}.
$$

To determine  $c_R$ independently,  we employ the method
introduced by one of the authors~\cite{P2013,P2014}.
We define the CTS function $Q_R$ as
\begin{align}
\label{Eq:QA}
Q_R(N;b) &:= \left \vert \log R_N + \ln R_{N/b^2} - 2 \ln R_{N/b} \right \vert \\
&= (b^{c_R}-1)^2 \left \vert B\right \vert N^{-c_R} + o(N^{-{c_R}}),
\nonumber
\end{align}
which indicates that the plots of $Q_R(N;b)/(b^{c_R}-1)^2$ against $N$ should converge onto a single curve, independent
of $b$, for large $N$, if the correct value of $c_R$ is used. By adjusting $c_R$ to achieve this
convergence in the asymptotic regime for various values of $b$, we can estimate $c_R$ without information about $-\beta/\bar \nu_c$.
This estimate can be further validated by measuring the slope in the log-log plot, as illustrated in the inset
of Fig.~\ref{Fig:nuH}. With this value of $c_R$, we can accurately estimate the FSS exponent $\bar \nu_c$, using
Eq.~\eqref{Eq:effective2}.

\section{\label{Sec:sh} Numerical estimate of $\bar \nu_c$ for the ES ($s=\frac12$) case}
In previous numerical studies~\cite{D1990,HCTP2015}, the FSS exponents for the ES case (regular sampling)
were suggested to be
$-{\beta}/{\bar \nu} \approx  -2/5$ and ${\gamma}/{\bar \nu} \approx 1/5$.
A notable feature of these estimates is their adherence to the hyperscaling  relation
in Eq.~\eqref{Eq:hyper}. Additionally, $\gamma=1/4$  was analytically predicted
in Ref.~\cite{D1990} as $K-K_c\rightarrow 0^+$.
However, the numerical estimates were obtained
from relatively small-scale data both in terms of system size and simulation time
($N=25~600$ and $t_\text{max}\approx 10^5$)~\cite{HCTP2015}, and are thus not fully convincing.
In our paper, the maximum system size ($N=2^{18}$) is approximately ten times larger
than that in the previous study~\cite{HCTP2015}.
To ensure the larger system reaches a stationary state and to collect sufficient stationary-state data,
we perform significantly longer simulations, extending up to $t_\text{max}\approx 10^9$, which is
approximately $10^4$ times longer.

\begin{figure}
\includegraphics[width=\linewidth]{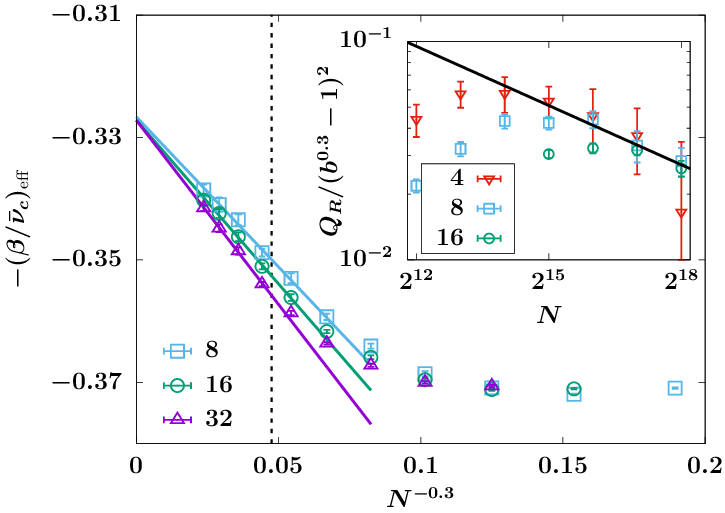}
\caption{\label{Fig:nuH}
Plots of $-(\beta/\bar \nu_c)_\text{eff}$ vs $N^{-0.3}$ for $s=\frac12$ with $b=8$ (square),
$16$ (circle), and 32 (triangle).
The straight lines on each dataset depict the results of the linear fitting.
The dashed vertical line indicates the position of the largest size ($N=25~600$) studied
in Ref.~\cite{HCTP2015}.
Inset: Double-logarithmic plots of $Q_R /(b^{0.3}-1)^2$ vs
$N$ for $b=4$ (reverse triangle), $8$ (square), and $16$ (circle).
The straight line with slope $-0.3$ is a guide for the eyes.
}
\end{figure}

Our numerical results are presented in Fig.~\ref{Fig:nuH}, where
the effective exponents $\xi^\textrm{eff}=-(\beta/\bar \nu_c)_\text{eff}$
are plotted against $N^{-c_R}$, with the estimated CTS exponent $c_R=0.3$.
In the inset, we display $Q_R /(b^{0.3}-1)^2$ plotted against $N$ for $b=4$, $8$, and $16$ on a double-logarithmic scale,
which demonstrates a tendency for the data collapse onto a straight line with slope $-0.3$
in the asymptotic regime of $N \gtrsim 2^{16}$. This behavior supports the choice of $c_R\approx 0.3$, independently
of $\beta/\bar \nu_c$.

For large $N$, Eq.~\eqref{Eq:effective2} predicts that the effective exponent $(\beta/\bar \nu_c)_\text{eff}$ approaches
the asymptotic value $\beta/\bar \nu_c$ linearly in terms of $N^{-c_R}$ with $b$-dependent slopes.
These predictions are consistent with numerical data shown in Fig.~\ref{Fig:nuH}.
Linear fitting for $N \ge 2^{15}$ yields
a consistent value of
\begin{align}\label{Eq:regular_FSS}
\beta/\bar \nu_c = 0.325 (15)\approx \frac13,~~\textrm{thus}~~ \bar \nu_c = 1.54 (7)\approx 
\frac32,
\end{align}
irrespective of the choice of $b$.
These values are  clearly distinct from $2/5$ and $5/4$, suggested in Ref.~\cite{HCTP2015}.

Notably, $(\beta/\bar \nu_c)_\text{eff}$ for small $N$
remains nearly constant around $0.37$, which is comparable to
the estimate $\beta/\bar \nu = 0.39(2)$ from Ref.~\cite{HCTP2015}.
However, it is evident that  $(\beta/\bar \nu_c)_\text{eff}$ eventually drifts toward a smaller value
approximately $0.325$, as $N$ increases.  This late crossover to the asymptotic scaling regime
renders the estimation of the FSS exponent exceedingly challenging. It also accounts for
the delayed onset of the power-law scaling of $Q_R$, as observed in the inset of Fig.~\ref{Fig:nuH}.

\section{\label{Sec:SC}Solutions of the self-consistent equation for entrained oscillators}
To pursue analytic understanding of the numerical results in the previous section,
this section investigates the MFSCE
for an arbitrary value of $s$.
As usual, we assume that the order parameter $\Delta_N(t)$ approaches a constant
value $R_N e^{i\psi_0}$ in the long-time limit (mean-field assumption). 
Then, the dynamic equations of Eq.~\eqref{Eq:kuraO} become completely decoupled
with unknown $R_N$
and their solutions are categorized into two distinct types.
Oscillators with $|\omega_k| < K R_N$ eventually stabilize at fixed angles $\theta^e_k$ (entrained oscillators),
satisfying  $\omega_k = K R_N \sin\theta_k^e$ (we set $\psi_0=0$ without loss of generality).
In contrast, oscillators with $|\omega_k| > K R_N$ continue to oscillate with modified frequencies (running oscillators).

By substituting these solutions into Eq.~\eqref{Eq:order0} and neglecting
the contribution from running oscillators, we arrive at the MFSCE 
\begin{align}\label{Eq:sce0}
R_N = \frac{1}{N} \sum_{k\in I^e} \sqrt{1 - \frac{\omega_k^2}{K^2 R_N^2}},
\end{align}
where $I^e$ denotes the index set of entrained oscillators ($|\omega_k| < K R_N$).
In this formulation, the contribution of running oscillators is disregarded to avoid
seemingly intractable complexity. Given that the contribution of running oscillators is 
at least $O(1/\sqrt{N})$ (representing independent random oscillators),
any solutions of the MFSCE at lower orders are meaningless. Thus, we seek solutions
only for $R_N\gg 1/N$ in the perturbation expansion.

For convenience, we define the following functions
\begin{align}
\label{Eq:Sz}
S(z)&:=
\frac{1}{N} \sum_{k \in I}
\Theta(z^2-\omega_k^2)
\sqrt{1 - \frac{\omega_{k}^2}{z^2} },\\
J(z)&:= \int_{-z}^z g(\omega)
\sqrt{1 - \frac{\omega^2}{z^2}} d \omega,
\end{align}
where $z>0$. It is clear that $J(z)=\lim_{N\rightarrow\infty} S(z)$.
With $g(\omega)$ as given in Eq.~\eqref{Eq:gden}, we obtain explicitly
$J(z)= \frac{3}{4}\pi z(1 - z^2)$.

Introducing the abbreviation $Z := K R_N$, we rewrite
the MFSCE, Eq.\eqref{Eq:sce0}, as
\begin{align}
\label{Eq:sc}
\frac{Z}K
=\frac{1}{N} \sum_{k}
\Theta(Z^2-\omega_k^2)
\sqrt{1 - \frac{\omega_{k}^2}{Z^2} }=S(Z),
\end{align}
which becomes, in the $N\rightarrow\infty$ limit,
\begin{align}
\label{Eq:Iz}
 \frac{Z}{K} = \frac{3}{4}\pi Z (1-Z^2).
\end{align}
The solution of Eq.~\eqref{Eq:Iz} is
\begin{align}
K_c = \frac{4}{3\pi},\quad
R= \begin{cases}
K^{-3/2}\sqrt{K-K_c}, & K \ge K_c,\\
0, &K<K_c,\end{cases}
\end{align}
which is consistent with $K_c = 2/[\pi g(0)]$ and $\beta = \frac12$.

We now examine the finite-size corrections, specifically how much $S(Z)$
deviates from  $J(Z)$ for large $N$. 
After a rather lengthy algebra (details are provided in the Appendix), we find
\begin{align}
\label{Eq:Scfin}
S(Z)\approx J(Z) + \frac{2s-1}{N}
+\frac{\eta_0(\alpha) }{\sqrt{g(0)Z N^3}},
\end{align}
where the $O(1)$ parameter $\alpha$ is defined as
\begin{align*}
\alpha:= N \left [G(Z)-x_{k_m} \right ], \quad k_m:={\rm max} \{k: \omega_k\le Z\},
\end{align*}
which represents the $N$-scaled distance in the $x_k$ space between the integral upper bound $Z$ and 
the largest natural frequency of the entrained oscillators $\omega_{k_m}$.
The coefficient $\eta_0$ is given as $\eta_0 = \eta+\eta_1$ with
\begin{align*}
&\eta(\alpha) = \sqrt{8\alpha} - \frac{2(1+2\alpha)^{3/2}}{3} + \frac{1}{12\sqrt{1+2\alpha}} \\
&\vert \eta_1(\alpha) \vert \le \tilde \eta(\alpha),\quad
\tilde \eta(\alpha):= \frac1{12}(1+2\alpha)^{-3/2}.
\end{align*}
Note that we have only bounds for $\eta_1$, which is actually enough for our purpose.

Now the MFSCE becomes
\begin{align}
\label{Eq:Scfina}
\frac{Z}{K} \approx \frac{Z}{K_c} ( 1 -  Z^2) + \frac{2s-1}{N}
+\frac{\eta_0(\alpha) }{\sqrt{g(0) Z N^3}}.
\end{align}
Note that the first correction term always dominates over the second one
for $Z\gg 1/N$ except for $s=\frac12$.

In deriving Eq.~\eqref{Eq:Scfin}, we have noticed
three sets of oscillators for which
the finite-size corrections exhibit different features:
(1) most of oscillators with equally spaced $x_k$'s,
(2) two oscillators in the edge region near $|\omega|\lesssim Z$, 
and (3) a few oscillators that break the equally spaced feature.
Set 1 contributes leading behavior $J(z)$ as well as the second order correction.
The second order correction also arises from set 2. 
The first order correction is from set 3; see also Sec.~\ref{Sec:samplingD} for further
discussion about set 3 in other schemes.

\begin{figure}
\includegraphics[width=\linewidth]{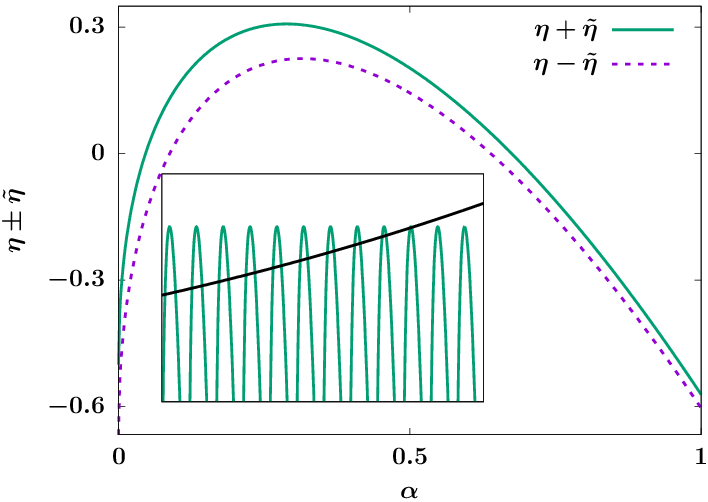}
\caption{\label{Fig:eta} Plots of
$\eta+\tilde \eta$ (top) and
$\eta- \tilde \eta$ (bottom) vs $\alpha$.
Since $-\tilde \eta(\alpha) \le \eta_1(\alpha) \le \tilde \eta(\alpha)$,
there should be a region where $\eta_0(\alpha)$ is positive.
Inset: Schematic graph to find solutions of Eq.~\eqref{Eq:ScfinH}.
The continuous curve depicts $y = x^{7/2}$.
}
\end{figure}
\subsection{ES case ($s=1/2$)}
We begin with analyzing the ES case ($s=\frac12$) at the critical point $K=K_c$,
which yields
\begin{align}
\label{Eq:ScfinH}
Z^{7/2} \approx K_c \frac{\eta_0(\alpha)}{\sqrt{g(0) N^3}}.
\end{align}
Since $\alpha$ is also a function of $Z$, we have to resort to a graphical method
to analyze Eq.~\eqref{Eq:ScfinH}.
In Fig.~\ref{Fig:eta}, we plot $\eta(\alpha)\pm \tilde \eta(\alpha)$ versus $\alpha$ that provide bounds
for $\eta_0$.
Since there is surely a region where $\eta_0$ is positive, Eq.~\eqref{Eq:ScfinH} must have a solution.

Observe that $\alpha$ is a (pseudo)periodic function of $Z$ in that 
$\alpha(Z+\delta Z)=\alpha(Z)$ if $N G(Z+\delta Z) - N G(Z) = 1$.
As $G(Z)\approx g(0)Z$ for small $Z$ in Eq.~\eqref{Eq:Gomega},
the period of $\alpha(Z)$ is approximately $1/[Ng(0)]$; 
see the inset of Fig.~\ref{Fig:eta}.
There are multiple solutions of Eq.~\eqref{Eq:ScfinH} [in fact, the number of solutions is $O(N^{4/7})$]
and the largest one should be $O(N^{-3/7})$. Although the selection of solutions
would depend on the initial condition in this framework, it may not worthy to discuss ``the'' solution
under the presence of noisy running oscillators. Nonetheless,
one may conclude that within the MFSCE the order parameter $R_N$ should be bounded from above 
(largest solution), thus
$\beta/\bar\nu_c \geq 3/7$ or $\bar\nu_c\leq 7/6\approx 1.17$. 

Note that this result is clearly inconsistent with our numerically obtained value
of $\beta/\bar\nu_c\approx 0.33$ or $\bar\nu_c\approx 1.54$ in Sec.~\ref{Sec:sh}
and also with the previous estimates of $\beta/\bar\nu=0.4$ or $\bar\nu=1.25$ in Ref.~\cite{HCTP2015}.
This implies that running oscillators cannot be neglected
and their contributions to $R_N$ may dominate the finite-size effects.

\subsection{$s\neq 1/2$}
For $s \neq \frac12$,  we rewrite the MFSCE, Eq.~\eqref{Eq:Scfin}, as
\begin{align}
\label{Eq:scO}
R_N \approx \frac{K}{K_c} R_N \left ( 1 - K^2 R_N^2 \right ) + \frac{2s-1}{N},
\end{align}
where the higher-order term is ignored for solutions with $R_N\gg 1/N$.
For $K>K_c$, one can easily find
\begin{align}
\label{Eq:nuSC}
R_N \approx \sqrt{\frac{K-K_c}{K^{3}}} \left ( 1
+ \frac{K_c K^{3/2}}{(K-K_c)^{3/2}}\frac{2s-1}{2N} \right ),
\end{align}
up to the leading order in the finite-size corrections.
In comparison to the scaling ansatz in Eq.~\eqref{Eq:Rchi_sf}, 
the MFSCE predicts $\bar \nu_+ =\frac32$ for $s\neq \frac12$.
In Sec.~\ref{Sec:sl}, we will argue, based on the simulation at criticality for $s>\frac12$, that
the MFSCE prediction is right for $s\neq \frac12$ in the supercritical region.
It is notable that numerically
obtained $\bar \nu_c$ for the ES
case  is very close to the perdiction of $\bar \nu_+$ by the MFSCE for $s \neq \frac12$.

The qualitative behavior of $R_N$ for $N \ll (K-K_c)^{-3/2}$ is quite different,
depending on the sign of $s-\frac12$.
When $s > \frac12$, the solution of $R_N$ in Eq.~\eqref{Eq:nuSC} is well defined for
any $N$. In particular, Eq.~\eqref{Eq:scO} for $K=K_c$ has a solution as
\begin{align}
\label{Eq:larges}
R_N \approx \left ( \frac{2s-1}{K_c^2} \right )^{1/3} N^{-1/3},
\end{align}
which predicts $\beta /\bar \nu_c = \frac13$, consistent with the conventional
scaling theory $\bar \nu_+ =\bar \nu_c$.
In this case, $R_N$ is nonzero even in the subcritical region ($K \lesssim K_c$), where
we get
\begin{align}
R_N \approx \frac{K_c}{K_c-K} \frac{2s-1}{N}.
\label{Eq:sub4}
\end{align}
Note that the higher-order term $1/\sqrt{R_N N^3} \sim \sqrt{K_c-K}/N$ is negligible in
comparison with $1/N$ in the regime ($|K-K_c| \ll 1$) we are interested in.
In comparison to the scaling ansatz in Eq.~\eqref{Eq:Rchi_sf}, 
the MFSCE also predicts $\bar \nu_- =\frac32$ for $s\neq \frac12$. That is, the MFSCE for $s>\frac12$
predicts a single FSS exponent just like the conventional theory.

\begin{figure}
\includegraphics[width=\linewidth]{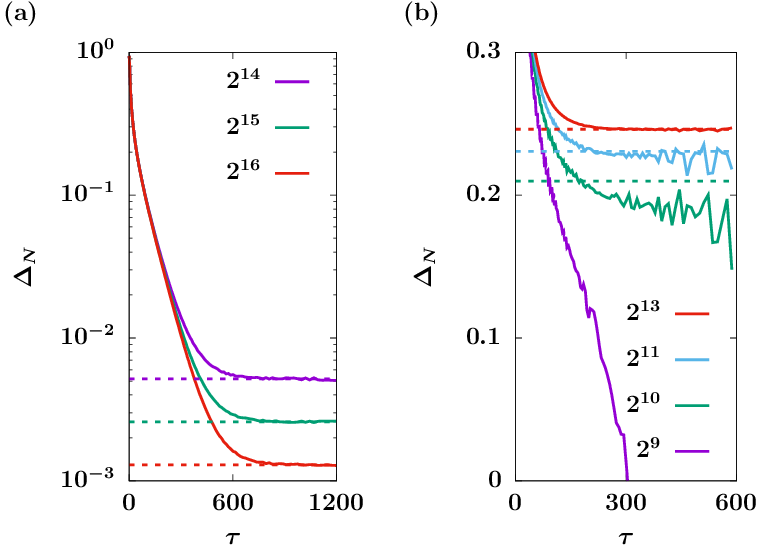}
\caption{\label{Fig:sup0}
(a) Semilogarithmic plots of $\Delta_N$ vs
$\tau$ for $s=1$ and $K=K_c-0.005$ (subcritical). The system sizes are $N=2^{14}$, $2^{15}$,
and $2^{16}$ (top to bottom).
Dashed horizontal lines indicate the values of Eq.~\eqref{Eq:sub4} for $N=2^{14}$, $2^{15}$,
and $2^{16}$ (top to bottom), which
show an excellent agreement with the saturating values of numerical integrations.
(b) Plots of $\Delta_N$ vs
$\tau$ for $s=0$ and $K-K_c =0.005$. The system sizes are $N=2^{9}$, $2^{10}$,
$2^{11}$, and $2^{13}$ (bottom to top).
Dashed horizontal lines indicate the values of Eq.~\eqref{Eq:nuSC} for
$N=2^{10}$, $2^{11}$, and $2^{13}$ (bottom to top). The predicted $R_N$
for $N=2^{13}$ shows an excellent agreement with the numerical data.
}
\end{figure}
In Fig.~\ref{Fig:sup0}(a), we compare the numerical integration of Eq.~\eqref{Eq:Rkura}
to the MFSCE prediction, Eq.~\eqref{Eq:sub4}, for $K = K_c - 0.005$ (subcritical) with $s=1$.
In contrast to the ES case,
even Eq.~\eqref{Eq:sub4} is in a perfect agreement with the numerical results
without any correction from running oscillators
up to the observed time in the figure.
If Eq.~\eqref{Eq:sub4} indeed prevails for all sufficiently large $N$ and for all time, 
the MFSCE would predict correct $\bar \nu_-$ for $s > \frac12$ and the conventional scaling theory (a single
FSS exponent $\bar \nu$)
would be applicable to this case.

Due to the condition $R_N\gtrsim N^{-1/2}$ to neglect running oscillators, however,
$K_c-K$ must be necessarily smaller than $N^{-1/2}$ for Eq.~\eqref{Eq:sub4} to be valid. 
Since the scaling ansatz should be applicable for all sufficiently large $N$ with $|K-K_c|$ 
small and fixed,
the running oscillators will eventually dominate as $N$ gets larger and, in turn, 
the prediction~\eqref{Eq:sub4} will eventually fail.
If this scenario is right, the seemingly successful prediction of the MFSCE in Fig.~\ref{Fig:sup0}(a)
can be argued to be due to the fact that the fluctuations by running oscillators are not fully developed
up to the observed time.

Once the contribution by running oscillators becomes dominant, the subcritical scaling function 
cannot be given by Eq.~\eqref{Eq:sub4}, for the size of $R_N$ is at least $O(1/\sqrt{N})$.
In fact, performing longer simulation than in Fig.~\ref{Fig:sup0}(a) revealed that
the seeming stationary behavior becomes disrupted at late time and 
$\Delta_N$ goes way down to negative, implying the running oscillators 
become dominant. Accordingly, $\bar \nu_-$ may be different from $\bar \nu_+$. 
This is clearly an interesting question in that the critical scaling of 
the Kuramoto model may not be described by the conventional scaling theory.
Since this question is beyond the scope of this paper,
we defer the full analysis of off-critical behavior to a later publication.

Let us move on to the case with $s<\frac12$.
At criticality and in the subcritical region ($K \le K_c$), there is no proper solution for $R_N\gg 1/N$.
Thus, running oscillators should be recalled to find any sensible value of $R_N$.
In other words, the MFSCE for $s<\frac12$ fails to predict $\bar \nu_c$ and $\bar \nu_-$.

Even in the supercritical region ($K>K_c$),
Eq.~\eqref{Eq:nuSC}
does not make sense for $N (K-K_c)^{3/2} \ll 1$ because the resulting $R_N$ becomes negative.
Thus, Eq.~\eqref{Eq:nuSC} is meaningful only for sufficiently large $N$ even in the
supercitical region.
In Fig.~\ref{Fig:sup0}(b), we present numerically integrated results for $s=0$ and
$K=K_c + 0.005$ for various $N$'s.
We observe that $\Delta_N$ for $N=2^{9}$ crosses
zero in short time even though $K > K_c$, implying an
easy escape from the fixed point for positive $\Delta_N$.
For $N\ge 2^{10}$,  $\Delta_N$ seems to saturate
to a nonzero value, which does not match with the value from Eq.~\eqref{Eq:nuSC} for $N \le 2^{11}$.
Only for $N=2^{13}$, the saturated value becomes consistent with the predicition
from Eq.~\eqref{Eq:nuSC}. Note that $N (K-K_c)^{3/2}\approx 2.90$ with $N=2^{13}$.

\section{\label{Sec:num} Numerical estimate of $\bar \nu_c$ for other cases ($s\neq \frac12$)}
This section analyzes numerical data at criticality
for various $s$, to find $\bar \nu_c$. 
We begin with presenting results for the case with $s>1/2$ in which the MFSCE seems working quite well.
\subsection{\label{Sec:sl}$s>1/2$}
\begin{figure}
\includegraphics[width=\linewidth]{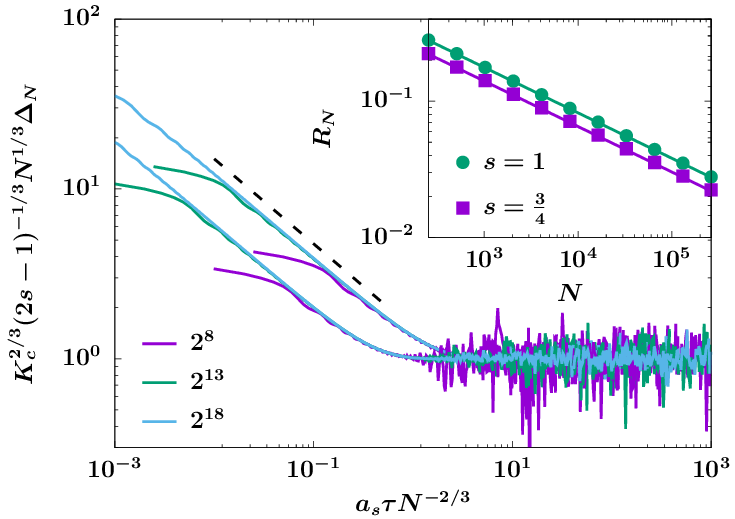}
\caption{\label{Fig:Lcol}Double logarithmic plots of $K_c^{2/3} (2s-1)^{-1/3} N^{1/3} \Delta_N$ vs
$a_s \tau  N^{-2/3}$ for $N=2^{8}$, $2^{13}$, $2^{18}$
and for $s=1$ (left three curves) and $s=\frac34$ (right three curves). 
For clean visibility, we set the different value of $a_s$ for different $s$
as $a_s = 0.4$ for $s=1$ and $a_s =1$ for $s=\frac34$.
The dashed line with slope $-0.5$ is a guide for the eyes.
Inset: Double logarithmic plots of $R_N$ vs $N$ for $s=1$ (solid circle) and
$s=\frac34$ (solid square). Error bars are smaller than the symbol size.
Corresponding straight lines depict $(2s-1)^{-1/3} K_c^{-2/3} N^{-1/3}$.}
\end{figure}

We first present the early-time behavior $\Delta_N$ ($\tau \le \tau_s$).
Recall that $\theta_k(0)  = 0$ for all $k$ and, therefore, $\Delta_N(0)=1$.
At the critical point for the infinite number of oscillators, we expect
\begin{align}
\label{Eq:Dmf}
\frac{dR}{d\tau} \propto - R^3,
\end{align}
which predicts $R(\tau) \sim \tau^{-1/2}$.
If $\Delta_N$ indeed approaches $R_N$ in Eq.~\eqref{Eq:larges},
there should be a typical time scale
$\tau_c$ such that $\tau_c^{-1/2} \sim N^{-1/3}$ or $\tau_c \sim N^{\bar z}$
with the {\em dynamic} exponent ${\bar z}=\frac23$.
If this is the case, a plot of $K_c^{2/3} (2s-1)^{-1/3}\Delta_N(\tau) N^{1/3}$
against $\tau / N^{2/3}$ would exhibit a data collapse, where the prefactor
in front of $\Delta_N$
is motivated from Eq.~\eqref{Eq:larges}.
To confirm this, we display the collapse plots in Fig.~\ref{Fig:Lcol}
for $s=\frac34$ and $1$, to observe perfect data collapses.
For a better visibility, we multiply an arbitrary factor in front of $\tau$ in Fig.~\ref{Fig:Lcol}.
It is remarkable that the saturating value in Fig.~\ref{Fig:Lcol} is almost 1,
consistent with the quantitative prediction in Eq.~\eqref{Eq:larges}.
This characteristic time scale is the motivation of choosing $N$ dependent $\tau_s = 10^3 N^{2/3}$ in Sec.~\ref{Sec:Meth}.

In the stationary state, the data become very noisy, which signals development of
large fluctuations by running oscillators.
We study $R_N$ by time averaging of $r_N(\tau)$ in the stationary state.
In the inset of Fig.~\ref{Fig:Lcol}, we plot $R_N$ against $N$ for $s=\frac34$
and $1$. For comparison, we also draw the result from Eq.~\eqref{Eq:larges},
which shows a nice coincidence. Since the prediction of Eq.~\eqref{Eq:larges} is so good,
we do not show the analysis of the effective exponent and conclude
$\bar \nu_c= \frac32$ for $s>\frac12$.

By our choice of the initial condition that gives $\theta_{-k}(t) = -\theta_k(t) \pmod{2\pi}$
for all $t$,
the system now has at most $Z_2$ symmetry (invariant only under $\theta_k \mapsto \pi + \theta_k$) rather than the continuous $U(1)$ symmetry.
Just like the Ising model equipped with dynamics (for instance, model A of Ref.~\cite{HH1977}),
it may still be possible for $\Delta_N$ to change its sign by fluctuations.
Within our simulation time, however,
we observed that $\Delta_N(\tau) = r_N(\tau)$ for all $\tau$ when $N \ge 2^{12}$.
In other words, $\Delta_N$ did not change its sign up to $\tau = \tau_s+\tau_\text{max}$ for
large $N$.
Even for smaller $N$, $\overline{\Delta_N}(\tau)$ is slightly different from $\overline{r_N}(\tau)$
(the size of difference is order of $10^{-6}$ for $N=32$).

Being deterministic, it is not at all
obvious whether this observation can be interpreted as a spontaneous $Z_2$ symmetry
breaking or that the dynamic equation never allows all the entrained oscillators to change
their phase by $\pi$. This question is clearly beyond the scope of our paper and we will not pursue
it any further.

Since $R_N$ in the supercritical region should be more stable than that at criticality,
stablility of the MFSCE solution at criticality (at least for long time) clearly suggests that 
in the supercritical region for $s\neq \frac12$ with
sufficiently large $N$, $R_N$ will be well approximated by the MFSCE and $\bar \nu_+$ is 
indeed $\frac32$ for any $s$ different from $\frac12$ (including small $s$). 
Our preliminary numerical investigation also supports this claim, which will be reported
elsewhere.

\begin{figure}
\includegraphics[width=\linewidth]{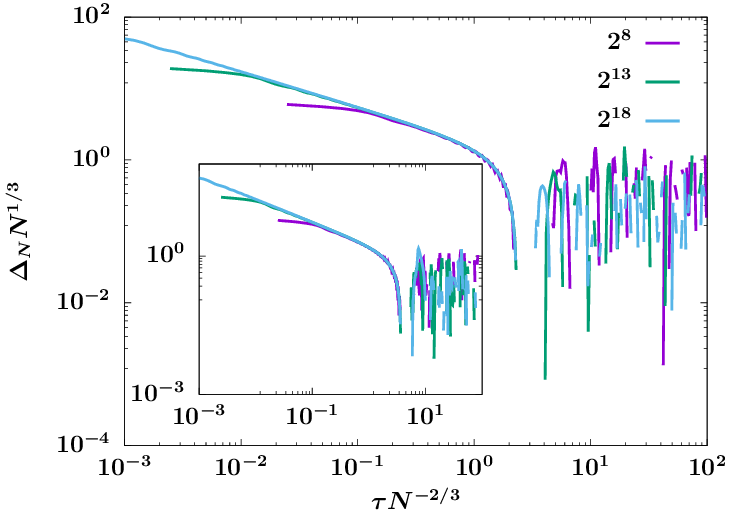}
\caption{\label{Fig:Scol} Double-logarithmic plots of $N^{1/3} \Delta_N$ vs
$\tau  N^{-2/3}$ for $s=0$  and  for $N=2^{8}$, $2^{13}$, and $2^{18}$.
All curves around $\tau N^{-2/3} \approx 1$ collapse well into a single curve.
The region without data points means $\Delta_N$ is negative.
Inset: Similar plots with the same number of oscillators for $s=\frac14$.
}
\end{figure}
\subsection{\label{Sec:ss}$s<\frac12$}
We begin with the early-time behavior.
In Fig.~\ref{Fig:Scol}, we plot $N^{1/3} \Delta_N$ against $\tau N^{-2/3}$
on a double logarithmic scale for $s=0$ (main figure) and for $s=\frac14$ (inset).
Absence of data points in some region simply means $\Delta_N$ is negative there.
All curves collapse perfectly into a single curve (for each $s$) before getting noisy.
Although the MFSCE at the critical point does not give a proper solution for this case,
$\bar \nu_+ = \frac32$ still plays the role in determining the typical
size and time scales of the system in the early-time regime.
The fact that $\Delta_N$ drops abruptly in an exponential way
combined with the absence of a proper solution in the MFSCE at criticality
suggests that there are, if they exist, only a few entrained oscillators and
running oscillators play a decisive role in the long-time behavior of 
the order parameters. 
We will further discuss the implication of this observation in Sec.~\ref{Sec:nuc}.

\begin{figure}
\includegraphics[width=\linewidth]{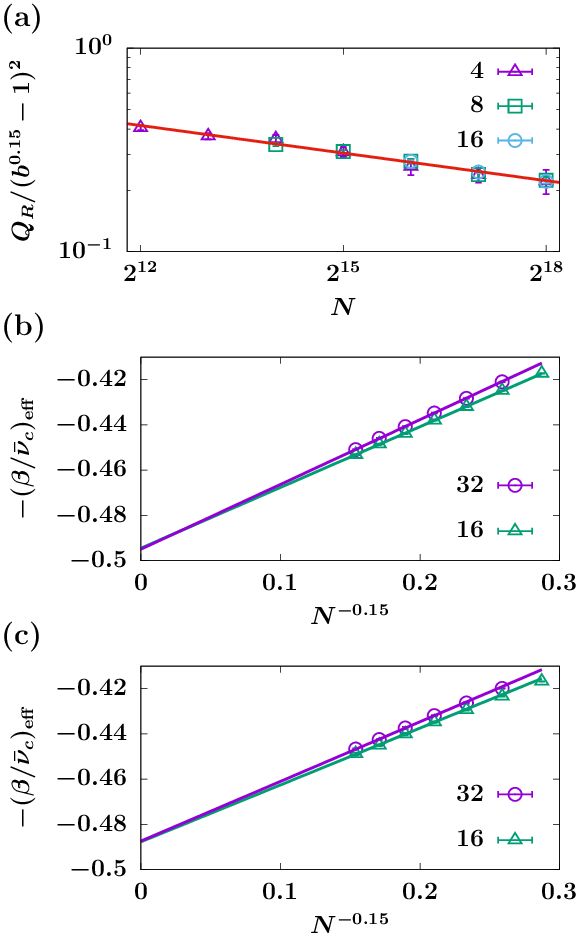}
\caption{\label{Fig:Sbn} (a) Double-logarithmic plots of $Q_R /(b^{0.15}-1)^2$ vs
$N$ with $b=4$, $8$, $16$ for $s=0$.
The straight line with slope $-0.15$ is a guide for the eyes.
(b) Plots of $-(\beta / \bar \nu_c)_\text{eff}$ vs $N^{-0.15}$ for $s=0$.
(c) Plots of $-(\beta / \bar \nu_c)_\text{eff}$ vs $N^{-0.15}$ for $s=\frac14$.
In (b) and (c), the straight lines show the linear fitting results of the corresponding
datasets.
}
\end{figure}
In Fig.~\ref{Fig:Sbn}(a), we depict $Q_R/(b^{0.15}-1)^2$ for $s=0$ as a function of $N$
for different values of $b$, to observe the same asymptotic behavior
for large $N$ regardless of $b$. We, therefore, conclude $c_R = 0.15$.
The $Q_R$ for $s=\frac14$ shows a similar power-law behavior to Fig.~\ref{Fig:Sbn}(a) and
we do not present details here.
With the obtained $c_R$,
we depict $-(\beta/\bar \nu_c)_\text{eff}$ as a function of $N^{-c_R}$ for
$s=0$ [Fig.~\ref{Fig:Sbn}(b)] and $s=\frac14$ [Fig.~\ref{Fig:Sbn}(c)], respectively,
for $b=16$ and $32$.
From the fitting, we conclude that $\beta /\bar \nu_c=0.49(1)$ or $\bar \nu_c
= 1.02(2)$ for both $s=0$ and $\frac14$,
which is obviously different from the supercritical $\bar \nu_+$.

The value $\beta /\bar \nu_c \approx 0.5$ is reminiscent of the order parameter
in case that all phases are iid random variables
with uniform distribution in the interval $(0,2\pi)$.
If the randomness is indeed the reason for the value $0.49(1)$, we would expect
$ \gamma = 0$. Our preliminary simulations, not unexpectedly, suggest that
$\gamma$ is nonzero. Hence, the behavior of $\Delta_N$ cannot be interpreted as
a fully random behavior of oscillators. Detailed analysis of fluctuations will be
reported elsewhere.

\subsection{\label{Sec:numsym} ES case ($s=\frac12$)}
We have already seen that $\bar \nu_c$ is not explained by the MFSCE for the ES case. 
By the study in the previous subsection, however, it is still worth asking whether
the anticipated upper bound of $\bar \nu_c$ from the MFSCE can play any role in the early time 
region at criticality.
Also it is natural to examine whether $\Delta_N$ saturates to a finite value (even
though ``the'' solution of the MFSCE cannot be fixed solely by the MFSCE) as in the case of $s> \frac12$
or decays exponentially just like the case of $s<\frac12$.

\begin{figure}
\includegraphics[width=\linewidth]{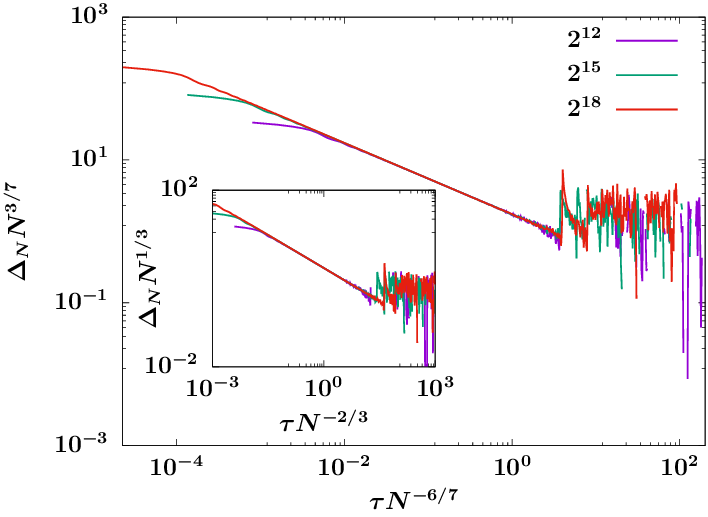}
\caption{\label{Fig:colH}
Plots of $\Delta_N N^{3/7}$ vs $\tau N^{-6/7}$ for $s=\frac12$ with $N=2^{12}$, $2^{15}$,
and $2^{18}$ on a double logarithmic scale.
Inset: Double-logarithmic plots of $\Delta_N N^{1/3}$ vs $\tau N^{-2/3}$ for the same data of the main figure.
}
\end{figure}
Since our simulation starts from $\Delta_N(0)=1$, it seems plausible to expect that the system would saturate
at the largest solution of the MFSCE and, therefore, $7/6$ of the MFSCE would play a role in the early
time regime.
Also by the same logic given in the beginning of Sec.~\ref{Sec:sl},
we would have the dynamic exponent to be $\bar z = 6/7$ obtained by equating $\tau_c^{-1/2} \sim N^{-3/7}$.
Based on this naive calculation, we depict $\Delta_N N^{3/7}$ as a function of $\tau N^{-6/7}$ 
in Fig.~\ref{Fig:colH} up to $\tau < \tau_s$.

Unlike the other cases, $\Delta_N$ shows neither saturating nor exponentially decaying behavior.
Rather, noisy behavior abruptly starts after exhibiting a power-law decay in time.
If we interpreted the beginning of the noisy behavior as the development of fluctuations,
this beginning time still seems to scale with the MFSCE dynamic exponent $\bar z = 6/7$.
For comparison, we plot $\Delta_N N^{1/3}$ against $\tau N^{-2/3}$
in the inset of Fig.~\ref{Fig:colH}. Although the collapse in the inset looks poorer than
the main figure, we, of course, cannot make a firm conclusion because of the lack
of either saturating or exponentially decaying behavior.

In a sense, the ES case seems to mix some features of other two cases.
First, similar to the case with $s>\frac12$, $\bar \nu_c$ is close to 1.5 as found in Sec.~\ref{Sec:sh}.
Second, similar to the case with $s<\frac12$, $\Delta_N$ fluctuates around zero for any $N$.

\section{\label{Sec:dis}Discussion}
\subsection{\label{Sec:samplingD} Sampling rule versus FSS}
Since our rule of choosing natural frequencies involves overall shift of $x_k$'s by the amount $s$
in comparison to the ES case,
it is natural to raise a question as to whether such a global shift of $x_k$'s is necessary to observe 
a different FSS. 
Also one may ask if any type of global shift always triggers a different FSS.
In fact, the answers are no for both questions as explained below. 

Let us answer the first question first.
One answer is already found by comparing the rule with $s=0$ and $1$ for even $N$. 
By the rule, $N-2$ oscillators have the same natural frequencies for these two cases.
To be more precise, we have $\omega_k (s=1) = \omega_{k-1} (s=0)$ for $2 \le k \le M$.
The only difference is that $\omega_{M}$ for $s=0$ is
replaced by $\omega_{1} = 0$ for $s=1$. 
That is, changing frequencies of two oscillators can trigger different FSS.
Since the entrainment is driven
by oscillators with small $\omega$, the change of scaling behavior
by assigning more oscillators around $\omega =0$ may not be surprising.
Still, we find it interesting for such a minute change to affect the global scaling behavior.

Presumably, this drastic change of scaling behavior by a minute modification 
should be attributed to 
the deterministic nature of dynamics as well as the fully connected interaction. 
It would be hardly conceivable to observe this effect if there is a ``thermal'' noise or if the system
is on a $d$-dimensional lattice.

We now give a more general answer, which is more insightful than the previous one.
We find it most effective to explain it using mathematical formulae.
Let 
$$
y_k := \frac{k-1/2}{N}, 
$$
for $k = 1,2,\ldots, N$ and
let $\Omega_0$ be the set of tuples $(k,y_k,\omega_k)$ with $\omega_k$ determined 
by the regular sampling rule Eq.~\eqref{Eq:regular}.
For convenience, the index set $I$ now consists of all positive integers from $1$ to $N$, 
which should not be confused with Sec.~\ref{Sec:sampling}.
Let
$$
S(z,\Omega_0) = \frac{1}N \sum_{k\in I} \Theta(z^2 - \omega_k^2) \sqrt{1 - \frac{\omega_k^2}{z^2}},
$$
which is just a repeat of Eq.~\eqref{Eq:Sz}.
For clarity, we explicitly write that $S$ is also a function of the set of tuples
$(k,y_k,\omega_k)$.
Now we consider two different sets $\Omega_\pm$ of tuples $(k,y^{\pm}_k,\omega^{\pm}_k)$ with
$$
y^{\pm}_k = \int_0^{\omega^{\pm}_k} g(\omega) d\omega.
$$
In $\Omega_+$, we assign $y^{+}_k = y_k$ for $k>1$ and choose $y^+_1$ such that 
$\omega^{+}_1 = a/N$ with $a$ an arbitrary $N$-independent constant. 
In $\Omega_-$, we assign $y^{-}_k = y_k$ for all $k \neq M$ ($M = \lfloor N/2\rfloor$: the integer part
of $N/2$) 
and choose $y^-_M$ such that $\omega^{-}_M = b$ with $b$ an arbitrary $N$-independent constant.
For later reference, we write $k_+=1$ and $k_-=M$.
Note that the full symmetry $\omega^{\pm}_k + \omega^{\pm}_{N-k+1} = 0$ is not 
satisfied in $\Omega_\pm$,
which actually does not play any role in the following discussion.

Assume $N$ is sufficiently large, $|z| \ll 1$, and $N z \gg 1$.
Let $\Delta_\pm S := 
S(z,\Omega_\pm) - S(z,\Omega_0)$.
Obviously,
\begin{align}
\label{Eq:Omega}
\Delta_\pm S =& \frac{1}{N} \Theta\left (z - \left \vert \omega_{k_\pm}^\pm \right \vert \right )
\sqrt{1 - \left ( \frac{\omega_{k_\pm}^\pm}{z} \right )^2}\\
&- \frac{1}{N} \Theta\left (z - \left \vert \omega_{k_\pm} \right \vert \right )
\sqrt{1 - \left ( \frac{\omega_{k_\pm}}{z} \right )^2}
\approx \pm \frac1N, 
\nonumber
\end{align}
regardless of the actual value of $a$ and $b$, where the error is  $O(z^{-2} N^{-3})$.
In other words, the MFSCE with $\Omega_+$ ($\Omega_-$) is identical to that of $s=1$ ($s=0$)
at least up to the second order correction. Again, a single change of 
a frequency can trigger
different FSS. We also performed simulations, to confirm the above conclusion at least for $\Omega_+$ 
(details not shown here).

In a more general setting, the above answer can be used to yield another MFSCE with 
different FSS exponents.
Assume $0 \le \mu < 1$ and we construct $\Omega_\mu$ such that
$y_{\mu,k} = a_k/N$ for $1 \le k \le N^\mu$ with $a_k$'s arbitrary $O(1)$ numbers
and $y_{\mu,k} = y_k$ for $k>N^\mu$.
Needless to say, $\Omega_\mu$ satisfies Eq.~\eqref{Eq:omega_cond}.
Then, by the same logic in Eq.~\eqref{Eq:Omega}, we have
$$
S(z,\Omega_\mu) \approx J(z) + N^{\mu-1},
$$
which obviously predicts $\mu$-dependent FSS exponents $\bar \nu_c = \bar \nu_+ = 1.5/(1-\mu)$. 
Note that in $S(z,\Omega_\mu) - S(z,\Omega_0)$, 
 $y_{\mu,k}$'s for $k \le N^\mu$ contribute $1/N$ for each, thus $N^{\mu-1}$.
Since more entrained oscillators are involved in the MFSCE as $\mu$ gets larger, which entails 
that the MFSCE becomes more and more reliable, and the MFSCE for the case of $\mu=0$ (or the case with $s=1$) 
already predicts the right $\bar \nu_+=\bar \nu_c$,  
we believe the MFSCE for $\Omega_\mu$ predicts correct $\bar \nu_+=\bar \nu_c$.
By preliminary numerical studies, we confirmed the above prediction for $\bar \nu_c$
with $\mu=1/4$ (not shown here). 

If $\mu >\frac12$, the MFSCE would predict the right $\bar \nu_-$ that is the same as $\bar \nu_+$,
because the solution [$\propto (K_c-K)^{-1}N^{\mu-1}$] of the MFSCE in the subcritical region is already larger than $N^{-1/2}$.
But numerical confirmation of this prediction for $\mu > \frac12$ would be very demanding,
since $N$ should be so large that 
Eq~\eqref{Eq:omega_cond} is well satisfied.

Now we move on to the second question. To this end, let us consider another rule of
regular sampling such that
\begin{align}
\label{Eq:Pazzo}
\int_{-\infty}^{\omega_k} g(\omega) d\omega = \frac{k}N,
\end{align}
which also globally shifts by $1/2$ from the ES case.
By noticing that the above rule can be implemented from the $s=0$ case
by reassigning $\omega_M$ to be zero with $\omega_{-M}$ intact, one can easily conclude that
the MFSCE with the rule~\eqref{Eq:Pazzo} becomes identical to $S(z,\Omega_0)$
up to the leading correction of $O(1/\sqrt{zN^3})$. In this sense, the above global shift 
does not affect the FSS in comparison to the ES case within the current context.
We also checked numerically that this conclusion is valid (details not shown here).
\subsection{\label{Sec:nuc} $\bar \nu_+$ versus $\bar \nu_c$ for $s=0$}
For $s=1$, we have argued (hopefully convincingly) that $\bar \nu_+ = \bar \nu_c$.
On the other hand, numerical simulations for $s=0$ at criticality showed that $\bar \nu_+\neq \bar \nu_c$.
According to the conventional scaling theory with $\rho_+(x)$, one would expect
for sufficiently small value of the scaling parameter $x$
\begin{align}
\label{Eq:smallx}
\rho_+(x) \sim x^{-\beta/\bar \nu_+},
\end{align}
which predicts $\bar \nu_+ = \bar \nu_c$.
Since $\rho_+(x)$ seems well defined even for $s=0$, one may ask how the scaling relation 
cannot be valid.

Actually, the answer is already given in Fig.~\ref{Fig:sup0}(b).
When $K-K_c$ is very small, $N$ should be 
sufficiently large for $\rho_+$ of the MFSCE 
to predict $\Delta_N$ correctly even qualitatively.
Therefore, no matter how large $N$ is, $x\rightarrow 0$ limit fails to have the
asymptotic behavior in Eq.~\eqref{Eq:smallx} and therefore $\bar \nu_+$ has no reason
to be associated with $\bar \nu_c$.
This should be compared with the case of $s=1$, where the MFSCE gives correct behavior 
for any $N$ at least qualitatively.
\begin{table}[b]
\caption{\label{Tab:nu}
The FSS exponents  
$\bar \nu_+$,
$\bar \nu_-$,
and, $\bar \nu_c$ estimated 
by the MFSCE and by simulations for different $s$.
X means the failure of the mean-field theory and 
a blank cell means not discussed in this paper.
The estimates of $\bar \nu_-$'s from simulations are not available in this paper, so we omit them in this table.
For comparison, the last column also presents the result of $\bar \nu$ in the literature for the ES case.}
\begin{ruledtabular}
\begin{tabular}{rcccccc}
 &\multicolumn{3}{c}{MFSCE}&\multicolumn{3}{c}{simulations}\\
\cline{2-4}
\cline{5-7}
 $s$&$\bar \nu_+$&$\bar \nu_-$&$\bar \nu_c$
 &$\bar \nu_+$&$\bar \nu_c$&$\bar \nu$\\
\hline
$=\frac12$&X&X&$\le \frac76$&&$\approx 1.52$&1.25\footnote{From Ref~\cite{HCTP2015}}\\
$>\frac12$&$\frac32$&$\frac32$&$\frac32$&$\frac32$&$\frac32$&\\
$<\frac12$&$\frac32$&X&X&$\frac32$&$\approx 1.02$&\\
\end{tabular}
\end{ruledtabular}
\end{table}

Similar discussion can be applied to the dynamic fluctuation, which
suggests that the fluctuation exponent $\gamma$ in the supercritical region may
not dictate the large-$N$ behavior of fluctuations at criticality. This is surely an intriguing question and a detailed discussion for $\gamma$ will be reported elsewhere.
\section{\label{Sec:sum}Summary and conclusion}
Up to now, we have analyzed the finite-size-scaling exponents 
$\bar \nu_\pm$ and $\bar \nu_c$ of the order parameter $R_N$ 
for the deterministic sampling of
natural frequencies with characterizing parameter $s$, 
using both the mean-field self-consistent equation and numerical simulations.
The results are summarized in Table~\ref{Tab:nu}.

The most significant contribution of this paper would be that by extensive simulations with
large size and long time we corrected the reported value in the literature of $\bar \nu$ 
of the regular-sampling case (the case with $s=\frac12$ in Table~\ref{Tab:nu}).
We have also shown that $\bar \nu_+$ need not be the same as $\bar \nu_c$ 
especially for the case with $s<\frac12$. We also argued that 
$\bar \nu_+$ in the supercritical region need not be identical to $\bar \nu_-$ 
in the subcritical region.
On this account, we claim that the finite-size scaling of the Kuramoto model should be 
reconsidered more carefully, especially for the deterministic sampling of natural
frequencies.
\begin{acknowledgments}
S.-C.P. acknowledges support by a National Research Foundation of Korea (NRF) grant funded by the 
Korea government (Grant No. RS-2023-00249949)
and The Catholic University of Korea, Research Fund, 2022. 
S.-C.P. also thanks the Korea Institute for Advanced Study (KIAS) for hospitality during his
sabbatical leave in 2023. 
H.P. acknowledges support by NRF Grant No. 2017R1D1A1B06035497
and individual KIAS Grant No. QP013601 at the KIAS.
The authors are grateful to the Center for Advanced Computation at 
KIAS for help with computing resources.
\end{acknowledgments}

\appendix*
\section{\uppercase{Derivation of finite-size corrections in the MFSCE}}
In this appendix, we derive Eq.~\eqref{Eq:Scfin} regardless of whether $N$ is even or odd.
To treat both even- and odd-$N$ cases on the same footing, 
we introduce $\sigma$ which takes 1 (0) if $N$ is odd (even) and write $x_k =
(k-s+\frac12 \sigma)/N$ .

We first define for  $0 \le x \le G(z)$
$$
f(x):= \sqrt{1 - \frac{G^{-1}(x)^2}{z^2}}.
$$
For brevity, $z$ is omitted in the argument of $f$.
Notice that by the change of variables $x=G(\omega)$, we have
$$
J(z)
=2 \int_0^{G(z)} f(x) dx.
$$
In the following, we assume $z \ll 1$ and $N z \gg 1$.
For later reference, we summarize the derivatives of $f$ for small $z$ as follows:
\begin{equation}
\label{Eq:fder}
\begin{aligned}
f(x) &\approx \frac{\sqrt{G(z)^2 - x^2}}{G(z)},\\
f'(x) &\approx -\frac{x}{G(z)\sqrt{G(z)^2- x^2}},\\
f''(x) &\approx - \frac{G(z)}{\left [ G(z)^2 - x^2 \right ]^{3/2}},
\end{aligned}
\end{equation}
where we have used $G^{-1}(x) \approx x/g_0$ and $G(z) \approx g_0 z$ with $g_0 := g(0)$.
Obviously, $f''(x)$ is a decreasing function of $x$ at least for sufficiently
small $z$.

It is convenient to define
\begin{align*}
k_m &:= \max\{k: \omega_k \le z\},\,
\alpha := NG(z) +s - \frac12 \sigma - k_m .
\end{align*}
Note that $\alpha$ is a function of $z$ with the range $0 \le \alpha \le 1$.
Let
\begin{align}
\nonumber
a^\pm_k&:= x_k \pm \frac1{2N},\quad z_0 := G^{-1}\left (a^-_{k_m} \right ),\\
\nonumber
J_0&:= 2 \int_0^{a^+_1} f(x) dx,\quad
J_1 := 2 \int_{G(z_0)}^{G(z)} f(x)dx,\\
J_2 &:= 2 \int_{a_2^-}^{G(z_0)} f(x) dx = 2 \sum_{k=2}^{k_m-1}
\int_{a^-_k}^{a^+_k} f(x) dx.
\label{Eq:Jdef}
\end{align}
Since $a^-_{k+1} = a^+_{k}$, we have $J(z) = J_0 + J_1 + J_2$.

We first find an approximation for $J_0$.
Since $f(x) = 1 + O(x^2/z^2)$ for $x \ll z$, we have
$$
J_0 
=2a_1^+ + O\left ( \frac1{z^2N^3} \right )
 = \frac{3-2s +\sigma}{N}+ O\left ( \frac1{z^2N^3} \right ).
$$
Since
$$
\frac1{N} \sum_{k=-1}^1 \sqrt{1-\frac{\omega_k^2}{z^2}} = \frac{2+\sigma}{N} + O\left ( \frac1{z^2N^3} \right ),
$$
we can write
\begin{align}
J_0 = \frac1{N} \sum_{k=-1}^1 \sqrt{1-\frac{\omega_k^2}{z^2}} + \frac{1-2s}{N}+ O\left ( \frac1{z^2N^3} \right ).
\label{Eq:J0}
\end{align}
Note that \eqref{Eq:J0} is valid for a general $g$ we are interested in.

If we write $g(\omega) = g_0(1 - 4 C \omega^2)$, we have
\begin{align*}
J_1
 =& z g_0(1-Cz^2)\cos^{-1}\left ( \frac{z_0}{z} \right ) \\
&+ g_0z_0\sqrt{1  - \frac{z_0^2}{z^2} } ( 2 C z_0^2 - Cz^2 - 1).
\end{align*}
Note that the above formula becomes an approximation for a general $g$.
Let $\epsilon := 1 - z_0/z$ or $z_0 = z(1-\epsilon)$.
Assuming $\epsilon \ll z \ll 1$, we can approximate $J_1 \approx g_0 4
\sqrt{2} z \epsilon^{3/2}/3 $.
Since $G(z) - G(z_0) = (1+2\alpha)/(2N)$ and $G(\omega) \approx g_0 \omega$
for small $\omega$, we further approximate
\begin{align}
\label{Eq:eps}
\epsilon =1-\frac{z_0}{z} \approx \frac{G(z) - G(z_0)}{G(z)} \approx
\frac{1+2\alpha}{2 g_0 z N},
\end{align}
which gives
\begin{equation}
\label{Eq:J1}
J_1 = \frac{2(1+2\alpha)^{3/2}}{3 }(g_0zN^3)^{-1/2}  +
o\left (\frac{1}{\sqrt{zN^3}}\right ).
\end{equation}
Note that the coefficient $C$ does not contribute to the leading order of \eqref{Eq:J1}.

To analyze $J_2$, we use the Taylor theorem.
For any $x$ with
$a_k^- \le x \le a_k^+$,
there is  $y_k$ such that
$a_k^- \le y_k \le a_k^+$ and
\begin{align}
f(x) =& f(x_k) + f'(x_k)(x-x_k)
+ \frac{f''(y_k)}{2}(x-x_k)^2,
\label{Eq:Tay}
\end{align}
where $y_k$ should be understood as a function of $x$.
Plugging \eqref{Eq:Tay} into \eqref{Eq:Jdef}, we have
\begin{align}
\label{Eq:J2}
J_2 &= \frac{2}{N} \sum_{k=2}^{k_m-1} f(x_k) + E_1,\\
\nonumber
E_1 &:=  \sum_{k=2}^{k_m-1} \int_{a_k^-}^{a_k^+} f''(y_k)
(x-x_k)^2 dx.
\end{align}
Since
$a_k^- \le y_k \le a_k^+$  and
$f''(x)$ is a decreasing function,
we have $f''(a_k^+) \le f(y_k) \le f(a_k^-)$ and, in turn,
$$
\frac{f''(a_k^+)}{12N^3}\le \int_{a_k^-}^{a_k^+} f''(y_k)(x-x_k)^2 dx \le \frac{f''(a_k^-)}{12N^3},
$$
where we have used
$$
\int_{a_k^-}^{a_k^+}(x-x_k)^2 dx
= \frac{1}{12N^3}.
$$
Therefore,
\begin{align*}
&E_2 -\frac{f''(a_2^-)}{12N^3} \le
E_1 \le E_2 -\frac{f''[G(z_0)]}{12N^3},\\
&E_2 := \frac1{12N^3}\sum_{k=2}^{k_m} f''(a_k^-).
\end{align*}
Since $f''(x)$ is monotonously decreasing and
$$
h(b)+\int_{a}^b h(k) dk \le \sum_{k=a}^b h(k) \le h(a) + \int_a^b h(k) dk
$$
for any monotonously decreasing function $h$ ($a \le b$ are integers), we have
\begin{align*}
\frac{f''[G(z_0)]}{12N^3} \le
E_2-\frac{f'[G(z_0)] - f'(a_2^-)}{12N^2} \le
\frac{f''[G(a_2^-)]}{12N^3}.
\end{align*}
Using \eqref{Eq:fder}, we have
\begin{align*}
\frac{f''[G(z_0)]}{12N^3} &= -\frac{(1+2\alpha)^{-3/2}}{12} \frac{1}{\sqrt{g_0z N^3}}
+ o\left (\frac{1}{\sqrt{zN^3}}\right ),\\
\frac{f'[G(z_0)]}{12N^2} &= -\frac{(1+2\alpha)^{-1/2}}{12} \frac{1}{\sqrt{g_0z N^3}}+ o\left (\frac{1}{\sqrt{zN^3}}\right ).
\end{align*}
Since the terms with $a_2^-$ are negligible, we have
\begin{align}
E_1 =& -
\left [ \frac{(1+2\alpha)^{-1/2} }{12} +\eta_1(\alpha)\right ]\frac{1}{\sqrt{g_0z N^3}} +
 o\left (\frac{1}{\sqrt{zN^3}}\right ),\nonumber \\
\label{Eq:E1}
&\vert \eta_1(\alpha) \vert \le \tilde \eta(\alpha),\quad
\tilde \eta(\alpha):= \frac1{12}(1+2\alpha)^{-3/2}.
\end{align}
Since $f^{(n)}[G(z_0)]/N^{n+1} = O(1/\sqrt{zN^3})$ for any $n$, it is difficult, if not impossible,
to find an explicit formula for $\eta_1$, even considering more terms in the Taylor expansion.

Since
$$
 \frac{2}{N}f(x_{k_m}) = \sqrt{8\alpha} (g_0zN^3)^{-1/2} + o\left ( \frac1{\sqrt{zN^3}} \right ),
$$
\eqref{Eq:J0}, \eqref{Eq:J1}, and \eqref{Eq:J2} combined with \eqref{Eq:E1} give Eq.~\eqref{Eq:Scfin}.
Now we have the approximate MFSCE for general $g$ as
\begin{align}
\frac{Z}{K} \approx \frac{Z}{K_c} ( 1 - C Z^2) + \frac{2s-1}{N}
+\frac{\eta_0(\alpha) }{\sqrt{g_0Z N^3}},
\end{align}
which is Eq.~\eqref{Eq:Scfina} for $C=1$ and $g_0=g(0)$.
Notice that $\eta_0$ is ``universal'' in that the above approximation is applicable
to a general $g$ with the corresponding $g_0$ and $C$.
By the same token, the conclusion based on the MFSCE in the main text is not limited to the
frequency density \eqref{Eq:gden}.

\bibliography{prebib,kura}
\end{document}